\documentclass[prl,aps,twocolumn,preprintnumbers,showpacs]{revtex4}     
\newcommand{\ket}[1]{|#1\rangle}    
\newcommand{\bra}[1]{\langle #1|} 
    
\begin{document}     
\title{Kinematic approach to the mixed state geometric phase in     
nonunitary evolution}     
\author{D.M. Tong$^1$, E. Sj\"{o}qvist$^2$\footnote{Electronic     
address: eriks@kvac.uu.se}, L.C. Kwek$^{1,3}$,     
and C.H. Oh$^1$\footnote{Electronic address: phyohch@nus.edu.sg}}     
\affiliation{$^1$Department of Physics, National University of     
Singapore, 10 Kent Ridge Crescent, Singapore 119260 \\     
$^2$Department of Quantum Chemistry, Uppsala University, Box 518,     
Se-751 20 Uppsala, Sweden \\     
$^3$National Institute of Education, Nanyang Technological     
University, 1 Nanyang Walk, Singapore 639798 }     
\date{\today}     
\begin{abstract}     
A kinematic approach to the geometric phase for mixed quantal states  
in nonunitary evolution is proposed. This phase is manifestly gauge  
invariant and can be experimentally tested in interferometry. It leads  
to well-known results when the evolution is unitary.  
\end{abstract}     
\maketitle     
\date{\today}     
The concept of geometric phase was first introduced by Pancharatnam    
\cite{Pancharatnam} in his study of interference of classical     
light in distinct states of polarization. Berry \cite{Berry}   
discovered the quantal counterpart of Pancharatnam's phase in the case   
of cyclic adiabatic evolution. Since then there has been an immense   
interest in holonomy effects in quantum mechanics, which has led to   
many generalizations of the notion of geometric phase. The extension   
to nonadiabatic cyclic evolution was developed by Aharonov and Anandan   
\cite{Aharonov}. Samuel and Bhandari \cite{Samuel} generalized    
the pure state geometric phase further by extending it to noncyclic
evolution and sequential projection measurements. The geometric phase
is a consequence of quantum kinematics and is thus independent of the
detailed nature of the dynamical origin of the path in state space.
This led Mukunda and Simon \cite{Mukunda} to put forward a kinematic
approach by taking the path traversed in state space as the primary
concept for the geometric phase. Further generalizations and
refinements, by relaxing the conditions of adiabaticity, unitarity,
and cyclicity of the evolution, have since been carried out
\cite{pati95}.
   
Another line of development has been towards extending the geometric    
phase to mixed states. This was first addressed by Uhlmann \cite{Uhlmann}    
within the mathematical context of purification.  Sj\"oqvist {\it et al.}   
\cite{Sjoqvistm} introduced an alternative definition of geometric   
phase for nondegenerate density operators based upon quantum  
interferometry.  Singh {\it et al.} \cite{Kuldip} gave a kinematic  
description of the mixed state geometric phase in  
Ref. \cite{Sjoqvistm} and extended it to degenerate  
density operators. The relation between phases of an entangled system  
and its subsystems has been investigated \cite{Sjoqvistp}. The concept  
of off-diagonal geometric phases in Ref. \cite{Manini} for pure states  
has also been generalized to mixed states undergoing unitary evolution  
\cite{filipp03}. Recently, the mixed state geometric phase   
in Ref. \cite{Sjoqvistm} has been verified experimentally using nuclear   
magnetic resonance technique   
\cite{du03}.    
     
The generalization of the mixed state geometric phase to nonunitary  
evolution has been addressed \cite{Ericsson03,Peixoto}. The concept  
proposed in Ref. \cite{Ericsson03} for completely positive maps (CPMs)  
is operationally well-defined but may yield different values of  
geometric phase for a given CPM when using different Kraus  
representations. The approach in \cite{Peixoto} also concerns the  
mixed state geometric phase for CPMs but is based upon a weaker form  
of parallel transport condition than \cite{Ericsson03}, which makes   
it unclear whether \cite{Peixoto} reduces to expected results   
\cite{Sjoqvistm,Kuldip} in the limit of unitary evolution. Related   
to these research efforts has been to analyze the effect of nonunitary  
processes on the pure state geometric phase \cite{garrison86,carollo03}.   
The lack of a clear consensus regarding the mixed state geometric   
phase in the nonunitary case, makes it important to pursue further   
studies on this issue.  
  
Geometric phases are useful in the context of quantum computing as a
tool to achieve fault tolerance \cite{zanardi99}. However, practical
implementations of quantum computing are always done in the presence
of decoherence. Thus, a proper generalization of the geometric phase
for unitary evolution to that for nonunitary evolution is central in
the evaluation of the robustness of geometric quantum computation. In
this Letter, we propose a quantum kinematic approach to the geometric
phase for mixed states in nonunitary evolution. We also propose a
scheme to realize nonunitary paths in the space of density operators
in the sense of purification, which could be of use in experimental
tests of the mixed state geometric phase.
  
Consider a quantum system $s$ with $N$ dimensional Hilbert space   
${\cal H}_s$. An evolution of the state of $s$ may be described   
as the path   
\begin{eqnarray}   
{\cal P}: t\in [0,\tau] \rightarrow \rho(t)=\sum\limits_{k=1}^N  
\omega_k(t) \ket{\phi_k (t)}\bra{\phi_k (t)},  
\label{rhot}   
\end{eqnarray}   
where $\omega_k(t) \geq 0$ and $\ket{\phi_k(t)}$ are the eigenvalues 
and eigenvectors, respectively, of the system's density operator
$\rho(t)$. All the nonzero $\omega_k(t)$ are assumed to be nondegenerate 
functions of $t\in [0,\tau]$, leaving the extension to the degenerate 
case to the end of the paper.  

To introduce the notion of mixed state geometric phase in nonunitary 
evolution, we begin by lifting the mixed state to a pure state in a 
larger system. Consider a combined system $s+a$ which consists of the 
considered system $s$ and an ancilla $a$ with $K \geq N$ 
dimensional Hilbert space. Without loss of generality, we assume  
in the following that $K=N$. The mixed state $\rho(t)$ can be lifted  
to the purified state 
\begin{eqnarray}     
\ket{\Psi(t)} = \sum\limits_{k=1}^N \sqrt{\omega_k(t)}     
\ket{\phi_k(t)}\otimes\ket{a_k} , \ t\in[0,\tau],  
\label{pt}  
\end{eqnarray}   
where $\ket{\Psi(t)}\in {\cal H}_s\otimes{\cal H}_a$ is a  
purification of the density operator of $s$ in the sense that  
$\rho(t)$ is the partial trace of $\ket{\Psi(t)}\bra{\Psi(t)}$  
over the ancilla. The Pancharatnam relative phase between 
$\ket{\Psi(\tau)}$ and $\ket{\Psi(0)}$ reads  
\begin{eqnarray}     
\alpha (\tau) & = & \arg\langle\Psi(0)|\Psi(\tau)\rangle 
\nonumber \\ 
 & = & \arg \left( \sum\limits_{k=1}^N  
\sqrt{\omega_k(0)\omega_k(\tau)}  
\bra{\phi_k(0)} \phi_k (\tau) \rangle \right) .  
\label{alphap}     
\end{eqnarray}     
Since both $\{\ket{\phi_k(0)} \}$ and $\{ \ket{\phi_k(t)} \}$ are 
orthonormal bases of the same Hilbert space ${\cal H}_s$, there 
exists, for each $t\in [0,\tau]$, a unitary  
operator $V(t)$ such that 
\begin{eqnarray}     
\ket{\phi_k(t)} = V(t) \ket{\phi_k(0)} ,   
\end{eqnarray}     
where $V(0)=I$, $I$ being the identity operator on ${\cal H}_s$.  
Explicitly, we may take 
\begin{eqnarray} 
V(t) = \ket{\phi_1 (t)} \bra{\phi_1 (0)} +  
\ldots + \ket{\phi_N (t)} \bra{\phi_N (0)}. 
\label{vt1} 
\end{eqnarray} 
Then, the relative  
phase can be recast as   
\begin{eqnarray}     
\alpha (\tau) =  
\arg \left( \sum\limits_{k=1}^N \sqrt{\omega_k(0)\omega_k(\tau)}  
\bra{\phi_k(0)} V(\tau)\ket{\phi_k(0)} \right) .    
\label{alphan}     
\end{eqnarray}     

In order to arrive at the geometric phase associated with the path 
${\cal P}$ of the state of $s$, we need to remove the dependence of 
$\alpha (\tau)$ upon the purification of the type displayed by Eq. 
(\ref{pt}). To do this, we first notice that $\alpha (\tau)$ becomes 
the standard geometric phase of the pure entangled state 
$\ket{\Psi(t)}, \ t\in[0,\tau]$ when the evolution satisfies the 
parallel transport condition $\langle \Psi(t) \ket{\dot {\Psi}(t)} = 
0$. However, this single condition is insufficient for mixed states as 
it only specifies one of the $N$ undetermined phases of $V(t)$, and 
the resulting pure state geometric phase remains strongly dependent 
upon the purification. Instead, the essential point to arrive at the 
geometric phase associated with ${\cal P}$ is to realize that there is 
an equivalence set ${\cal S}$ of unitarities $\widetilde{V}(t)$ that 
for $t\in [0,\tau]$ all realize ${\cal P}$, namely those of the form 
\begin{eqnarray}     
\widetilde{V}(t) =     
V(t) \sum\limits_{k=1}^N e^{i\theta_k(t)}  
\ket{\phi_k(0)} \bra{\phi_k(0)} ,     
\label{uuea2}     
\end{eqnarray}     
where $V(t) \in {\cal S}$ fulfills $V(0)=I$, but is otherwise 
arbitrary and $\theta_k (t)$ are real time-dependent parameters such 
that $\theta_k(0)=0$.  
We may in particular identify $V^{\parallel} (t) \in {\cal S}$  
fulfilling the parallel transport conditions 
\begin{eqnarray}     
\bra{\phi_k (0)} V^{\parallel\dagger} (t)     
\dot{V}^{\parallel} (t) \ket{\phi_k (0)} = 0 , \ k=1,\ldots,N .     
\label{parallel}     
\end{eqnarray}     
in terms of which the relative phase in 
Eq. (\ref{alphan}) coincides with the geometric phase associated with 
the path ${\cal P}$. Substituting $V^{\parallel} (t) = \widetilde{V} 
(t)$, with $\widetilde{V} (t)$ given by Eq. (\ref{uuea2}), into Eq. 
(\ref{parallel}), we obtain 
\begin{eqnarray}    
\theta_k (t) = i \int_0^t \bra{\phi_k(0)} V^{\dagger} (t')    
\dot V (t') \ket{\phi_k(0)} dt'.     
\end{eqnarray}    
Taking this expression for $\theta_k(t)$ into Eq. (\ref{alphan})     
for $V^{\parallel} (t)$, we finally obtain the geometric phase for     
the path ${\cal P}$ as    
\begin{eqnarray}     
\gamma [{\cal P}] = \arg \left( \sum\limits_{k=1}^N     
\sqrt {\omega_k(0) \omega_k(\tau)} \bra{\phi_k(0)} V^\parallel(\tau)   
\ket{\phi_k(0)} \right).   
\label{gamman1} 
\end{eqnarray}     
The explicit expression of it reads 
\begin{widetext}     
\begin{eqnarray}     
\gamma [{\cal P}] =\arg \left( \sum\limits_{k=1}^N     
\sqrt{\omega_k(0)\omega_k(\tau)}\langle \phi_k(0)|\phi_k(\tau)\rangle     
e^{-\int_0^\tau \langle \phi_k(t)|\dot \phi_k(t) \rangle dt} \right).     
\label{gamman}     
\end{eqnarray}     
\end{widetext}     
   
Now, a reasonable notion of mixed state geometric phase in the  
nonunitary case should satisfy the conditions: a) it must be gauge  
invariant, i.e., only be dependent upon the path traced out by the  
system's density operator $\rho(t)$; b) it should reduce to well-known  
results in the limit of unitary evolution; c) it should be  
experimentally testable. Let us verify that geometric phase in  
Eq. (\ref{gamman}) fulfills these conditions. 
   
First, the phase $\gamma [{\cal P}]$ is manifestly gauge invariant 
in that it takes the same value for all $V(t) \in {\cal S}$. One may 
check this point by directly substituting Eq. (\ref{uuea2}) into 
Eq. (\ref{gamman}) and find 
\begin{eqnarray}     
\gamma [{\cal P}] \big|_{V(t)} =    
\gamma [{\cal P}] \big|_{\widetilde{V}(t)} .     
\label{gammauu}     
\end{eqnarray}     
In particular, if we let $V(t)=V^{\parallel}(t)$, we have     
\begin{eqnarray}     
\gamma [{\cal P}] & = & \arg \left( \sum\limits_{k=1}^N     
\sqrt{\omega_k(0)\omega_k(\tau)} \bra{\phi_k(0)}     
V^{\parallel} (\tau) \ket{\phi_k(0)} \right)
\nonumber \\ 
 & = & \alpha (\tau) ,  
\label{gammans}     
\end{eqnarray}     
which verifies that the relative phase gives the geometric phase   
for $V(t)=V^{\parallel} (t)$. Thus, the geometric phase defined  
by Eq. (\ref{gamman}) depends only upon the path ${\cal P}$ traced  
out by $\rho(t)$.  
   
Secondly, when the evolution is unitary, corresponding to the case 
where the eigenvalues $\omega_k$ are time independent and $V(t)$ is 
identified with the time evolution operator of the state, the 
geometric phase defined by Eq. (\ref{gamman}) leads to well-known 
results \cite{Sjoqvistm,Kuldip} .   
 
Finally, we demonstrate that the phases $\alpha (\tau)$ and $\gamma 
[{\cal P}]$ are experimentally testable. The measurement can be done 
by using the scheme of purifying $\rho(t)$ described in 
Eq. (\ref{pt}). In fact, the interference profile between 
$|\Psi(0)\rangle$ and $|\Psi(\tau)\rangle$ reads 
\begin{eqnarray}     
{\cal I} (\chi) & = & 
\left| e^{i\chi} \ket{\Psi(0)} + \ket{\Psi(\tau)} \right|^2      
\nonumber \\ 
 & \propto & 1+\nu(\tau)\cos[\chi-\alpha (\tau)] ,     
\label{i}     
\end{eqnarray}     
where $\alpha(\tau)$ is the relative phase in Eq. (\ref{alphan}), 
and     
\begin{equation}    
\nu(\tau) = \Big| \sum\limits_{k=1}^N    
\sqrt{\omega_k(0)\omega_k(\tau)} \bra{\phi_k(0)}   
\phi_k(\tau) \rangle \Big|     
\end{equation}     
is the visibility of the interference fringes obtained by varying the 
additional U(1) shift $\chi$. Using the Mach-Zehnder interferometer 
setup with $\ket{\Psi(0)}$ and $\ket{\Psi(\tau)}$ as internal states  
in each beam, the intensity modulation can be measured and the phase 
$\alpha(\tau)$ is obtained.   
  
A construction of the purification Eq. (\ref{pt}) of the path ${\cal P}$  
is as follows. Let $U_{sa}(t)$ be a unitarity on ${\cal H}_s \otimes  
{\cal H}_a$ such that $\ket{\Psi(t)} = U_{sa}(t) \ket{\Psi(0)}, \  
t\in[0,\tau]$, purifies the path ${\cal P} : t\rightarrow \rho(t)$.  
The desired purifications are obtained for all choices of $U_{sa} (t)$  
for which the ancilla part of the Schmidt basis of the tensor product  
space ${\cal H}_s \otimes {\cal H}_a$ is kept fixed. Explicitly,  
$U_{sa}(t)$ may be expressed as 
\begin{eqnarray}     
U_{sa}(t) = (V (t) \otimes I) W(t) W^{\dagger} (0),    
\label{uia}     
\end{eqnarray}     
where $\{ W(t) | t\in [0,\tau] \}$ is a one-parameter family   
of unitary operators on ${\cal H}_s \otimes {\cal H}_a$. These   
latter operators are restricted only by the requirement that the   
elements of the $k_0l_0$-th column, say, of their matrix representation   
in the $|\phi_k(0) \rangle \otimes |a_l\rangle$ basis must obey      
\begin{eqnarray}   
W_{kl,k_0 l_0} (t) = \delta_{kl} \sqrt{\omega_k (t)},   
\label{omega}     
\end{eqnarray}     
where $k,l,k_0,l_0=1,\ldots,N$.  With $U_{sa}(t)$, the  
relative phase $\alpha (\tau)$ is measured via Eq. (\ref{i}) and   
it gives the geometric phase either if $V(t)=V^\parallel(t)$ or  
exposing the other beam by a compensating unitarity of the form  
\begin{equation} 
V_c (t) = \sum_{k=1}^N e^{\int_0^{t} \bra{\phi_k(0)} V^{\dagger} (t')  
\dot{V} (t') \ket{\phi_k (0)} dt'} \ket{\phi_k (0)}\bra{\phi_k (0)}   
\end{equation} 
resulting in the relative unitarity $V(t) V_c^{\dagger}(t) = 
V^{\parallel} (t)$, acting on $s$. Thus, we have demonstrated 
that the present mixed state geometric phase is experimentally 
testable in principle \cite{example}.   
  
To calculate the geometric phase for an explicit physical example,   
let us consider a qubit subjected to the free precession Hamiltonian   
$H = (\eta /2) \sigma_z$ and dephasing represented by the Lindblad   
operator \cite{lindblad76} $\Gamma = \sqrt{(\Lambda /2)} \, \sigma_z$,   
where the real parameters $\eta$ and $\Lambda$ are the precession   
rate and strength of dephasing, respectively. For the qubit initially   
in a pure state characterized by the Bloch vector   
${\bf r} (0) = (\sin \theta_0,0,\cos \theta_0)$, the solution   
$\rho_{\textrm{dp}} (t)$ of the Lindblad equation  
\cite{lindblad76} is characterized by     
\begin{eqnarray}   
\omega_1 (t) & = & 1-\omega_{2} (t) 
\nonumber \\ 
 & = & \frac{1}{2} 
\left( 1 + \sqrt{\cos^2 \theta_0 +   
e^{-2\Lambda t} \sin^2 \theta_0} \right) ,   
\nonumber \\   
\ket{\phi_1 (t)} & = & e^{-i\eta t/2} \cos \frac{\theta_t}{2} \ket{0} +   
\sin \frac{\theta_t}{2} e^{i\eta t/2} \ket{1} ,  
\nonumber \\   
\ket{\phi_2 (t)} & = & -e^{-i\eta t/2} \sin \frac{\theta_t}{2} \ket{0} +   
\cos \frac{\theta_t}{2} e^{i\eta t/2} \ket{1} ,   
\label{lindsol1}  
\end{eqnarray}  
where   
\begin{eqnarray}   
\tan \theta_t = e^{-\Lambda t} \tan \theta_0    
\label{lindsol2}  
\end{eqnarray}  
and $\{ \ket{0},\ket{1}\}$ is the standard qubit basis. By inserting 
Eqs. (\ref{lindsol1}) and (\ref{lindsol2}) into Eq. (\ref{gamman}), 
the geometric phase associated with the quasi-cyclic path ${\cal P} : 
t \in [0,2\pi/\eta] \rightarrow 
\rho_{\textrm{dp}} (t)$ becomes (assuming $\cos \theta_0 \geq 0$)  
\begin{widetext}   
\begin{eqnarray}   
\gamma [{\cal P}] = -\pi + \frac{\eta}{4\Lambda}   
\ln \left( \frac{\big( 1-\cos \theta_0 \big)   
\big( \sqrt{\cos^2 \theta_0 + \sin^2 \theta_0 \,   
e^{-4\pi \Lambda /\eta}} +   
\cos \theta_0 \big)}{\big( 1+\cos \theta_0 \big)   
\big( \sqrt{\cos^2 \theta_0 + \sin^2 \theta_0 \,   
e^{-4\pi \Lambda /\eta}} -    
\cos \theta_0 \big)} \right) .   
\label{dpgp}  
\end{eqnarray}  
\end{widetext}  
For small $\Lambda / \eta$, we may Taylor expand the right-hand   
side of Eq. (\ref{dpgp}) and obtain to first order   
\begin{eqnarray}   
\gamma [{\cal P}] \approx -\pi (1-\cos \theta_0) + \pi^2 \cos \theta_0  
\sin^2 \theta_0 \frac{\Lambda}{\eta} .  
\end{eqnarray}   
In Ref. \cite{carollo03}, the effect of dephasing on the pure state  
geometric phase has been analyzed using a quantum-jump approach,  
leading to a dephasing independent geometric phase effect. From  
the perspective of the mixed state geometric phase, we have obtained  
a first order dependence on dephasing which only reduces to that of  
Ref. \cite{carollo03} for nonunitary paths ${\cal P}$ characterized   
by $\theta_0 = \pi/2$, corresponding to precession in the equatorial  
plane of the Bloch ball.  

Let us end by briefly delineating the degenerate case. 
Consider the path
\begin{eqnarray} 
{\cal P}: t\in [0,\tau] \rightarrow \rho(t) = 
\sum_{k=1}^K \sum_{\mu=1}^{n_k} \omega_k(t) 
\ket{\phi_{k}^{\mu} (t)} \bra{\phi_{k}^{\mu} (t)},
\end{eqnarray}  
where $\omega_k(t)$, $k=1,\ldots,K\leq N$, are the eigenvalues 
of $\rho(t)$ each with degeneracy $n_k$, and $\ket{\phi_{k}^{\mu} (t)}$, 
$\mu=1,\ldots,n_k$, are the corresponding degenerate eigenvectors. 
The geometric phase of ${\cal P}$ is 
\begin{widetext}
\begin{eqnarray}     
\gamma [{\cal P}] & = & \arg \left( \sum_{k=1}^K 
\sum_{\mu=1}^{n_k} \sqrt {\omega_k(0) \omega_k(\tau)} 
\bra{\phi_{k}^{\mu}(0)} V^\parallel(\tau) \ket{\phi_{k}^{\mu}(0)} \right). 
\label{gamman2} 
\end{eqnarray} 
\end{widetext}
In the above expression, $V^\parallel(\tau)$ is defined by
$V^\parallel(t)=V(t)\sum_k V_k (t)$ with $V(t)=\sum_{k,\mu} 
\ket{\phi_{k}^{\mu} (t)}\bra{\phi_{k}^{\mu} (0)}$ and
\begin{eqnarray} 
V_k(t)= 
\sum_{\mu,\nu} \ket{\phi_{k}^{\mu} (0)} \bra{\phi_{k}^{\nu} (0)} 
\alpha_k^{\mu\nu} (t), 
\end{eqnarray} 
where $\alpha_k^{\mu\nu} (t)$ are determined by the parallel 
transport condition 
\begin{eqnarray} 
\bra{\phi_{k}^{\mu} (0)} V^{\parallel\dagger} (t)      
\dot{V}^{\parallel} (t) \ket{\phi_{k}^{\nu} (0)} 
 & = & 0 \ \mu,\nu =1, \ldots, n_k 
\label{parallel2}      
\end{eqnarray} 
with $\alpha_k(0)=I$, which leads to
\begin{eqnarray} 
\alpha_k^{\mu\nu} (t) = \bra{\phi_{k}^{\mu} (0)} {\textrm{P}} 
e^{-\int_{0}^{t} V(t')^{\dag}{\dot V}(t')dt'} 
\ket{\phi_{k}^{\nu} (0)} , 
\label{alphak}   
\end{eqnarray} 
where P denotes path ordering. The above may be generalized to the 
case where $\omega_k (t)$ is degenerate only on the time interval 
$[t_0,t_1] \subset [0,\tau]$ by noting that the eigenvectors in the 
corresponding subspace are, due to continuity, uniquely given at the 
end points $t=t_0$ and $t=t_1$.  

In summary, we have proposed a kinematic approach to the mixed state
geometric phase in nonunitary evolution. The proposed geometric phase
is gauge invariant in that it only depends upon the path in state
space of the considered system. We have demonstrated that the proposed
geometric phase for nonunitarily evolving mixed states is
experimentally testable in interferometry. Moreover, it leads to the
well-known results when the evolution is unitary. 
As an example, we have used the present
approach to calculate the geometric phase for nonunitarily evolving
mixed states in the case of a qubit undergoing free precession around
a fixed axis and affected by dephasing. 
\vskip 0.3 cm   
The work by Tong was supported by NUS Research Grant No. 
R-144-000-071-305. E.S. acknowledges financial support from the 
Swedish Research Council. 
     

\begin{thebibliography}{99} 
\bibitem{Pancharatnam} S. Pancharatnam,  
Proc. Indian Acad. Sci., Sect. A {\bf 44}, 247 (1956).     
\bibitem{Berry} M.V. Berry,     
Proc. R. Soc. London Ser. A {\bf 392}, 45 (1984).     
\bibitem{Aharonov} Y. Aharonov and J. Anandan,     
Phys. Rev. Lett. {\bf 58}, 1593 (1987);   
J. Anandan and Y. Aharonov,     
Phys. Rev. D {\bf 38}, 1863 (1988).     
\bibitem{Samuel} J. Samuel and R. Bhandari,     
Phys. Rev. Lett. {\bf 60}, 2339 (1988).     
\bibitem{Mukunda} N. Mukunda and R. Simon,     
Ann. Phys. (N.Y.) {\bf 228}, 205 (1993).     
\bibitem{pati95} A.K. Pati,     
Phys. Rev. A {\bf 52}, 2576 (1995); 
J. Phys. A {\bf 28}, 2087 (1995).     
\bibitem{Uhlmann} A. Uhlmann,     
Rep. Math. Phys. {\bf 24}, 229 (1986);     
Lett. Math. Phys.{\bf 21}, 229 (1991).     
\bibitem{Sjoqvistm} E. Sj\"oqvist {\it et al.}, 
Phys. Rev. Lett. {\bf 85}, 2845 (2000).     
\bibitem{Kuldip} K. Singh {\it et al.}, 
Phys. Rev. A  {\bf 67}, 032106 (2003).     
\bibitem{Sjoqvistp} E. Sj\"oqvist,     
Phys. Rev. A {\bf 62}, 022109 (2000);    
B. Hessmo and E. Sj\"oqvist,     
Phys. Rev. A {\bf 62}, 062301 (2000);     
D.M. Tong {\it et al.}, 
J. Phys. A {\bf 36}, 1149 (2003);   
D.M. Tong {\it et al.},      
Phys. Rev. A, {\bf 68}, 022106 (2003);   
M. Ericsson {\it et al.}, 
Phys. Rev. Lett. {\bf 91}, 090405 (2003).     
\bibitem{Manini} N. Manini and F. Pistolesi,     
Phys. Rev. Lett. {\bf 85}, 3067 (2000).     
\bibitem{filipp03} S. Filipp and E. Sj\"{o}qvist,     
Phys. Rev. Lett. {\bf 90}, 050403 (2003);       
Phys. Rev. A {\bf 68}, 042112 (2003).     
\bibitem{du03} J.F. Du {\it et al.}, 
Phys. Rev. Lett. {\bf 91}, 100403 (2003).     
\bibitem{Ericsson03} M. Ericsson {\it et al.}, 
Phys. Rev. A {\bf 67}, 020101(R) (2003).     
\bibitem{Peixoto} J.G. Peixoto de Faria {\it et al.}, 
Europhys. Lett. {\bf 62}, 782 (2003).     
\bibitem{garrison86} J.C. Garrison and E. Wright,    
Phys. Lett. A {\bf 128}, 177 (1986);    
D. Ellinas {\it et al.},  
Phys. Rev. A {\bf 39}, 3228 (1989);   
K.M. Fonseca Romero {\it et al.},     
Physica A {\bf 307}, 142 (2002);   
A. Nazir {\it et al.},     
Phys. Rev. A {\bf 65}, 042303 (2002);   
R.S. Whitney and Y. Gefen,    
Phys. Rev. Lett. {\bf 90}, 190402 (2003);   
G. De Chiara and G.M. Palma,    
Phys. Rev. Lett. {\bf 91}, 090404 (2003);     
A. Carollo {\it et al.},      
Phys. Rev. Lett. {\bf 92}, 020402 (2004). 
\bibitem{carollo03} A. Carollo {\it et al.}, 
Phys. Rev. Lett. {\bf 90}, 160402 (2003). 
\bibitem{zanardi99} P. Zanardi and M. Rasetti,   
Phys. Lett. A {\bf 264}, 94 (1999);   
J.A. Jones {\it et al.},    
Nature (London) {\bf 403}, 869 (2000);      
A.K. Ekert {\it et al.}, 
J. Mod. Opt. {\bf 47}, 2501 (2000);      
G. Falci {\it et al.}, 
Nature {\bf 407}, 355 (2000);   
L.-M. Duan {\it et al.}, 
Science {\bf 292}, 1695 (2001);   
X.B. Wang and Matsumoto Keiji,     
Phys. Rev. Lett. {\bf 87}, 097901 (2001).     
\bibitem{example} As an example of the operators $W(t)$ and 
$U_{sa}(t)$, suppose that the considered system and ancilla 
are qubit (two-level) systems. In this case, $W(t)$ may be 
chosen as 
\begin{eqnarray}     
 & W(t) & = \sqrt{\omega_1(t)}\sigma_z\otimes  
I+\sqrt{\omega_2(t)}\sigma_x\otimes \sigma_x ,  
\nonumber     
\end{eqnarray}     
which is unitary and fulfills Eq. (\ref{omega}). It yields   
\begin{eqnarray}     
U_{sa}(t) = (V(t)\otimes I) (\zeta I \otimes I +   
i\xi \sigma_y \otimes \sigma_x) .   
\nonumber   
\end{eqnarray} 
Here, $\sigma_x, \sigma_y ,\sigma_z$ are the standard Pauli   
operators and    
\begin{eqnarray}    
\zeta & = & \sqrt{\omega_1(0)\omega_1(t)} +     
\sqrt{\omega_2(0)\omega_2(t)} ,     
\nonumber \\     
\xi & = & \sqrt{\omega_1(t)\omega_2(0)}-  
\sqrt{\omega_2(t)\omega_1(0)}.   
\nonumber  
\end{eqnarray}     
There are infinitely many other choices of $W(t)$, and thus 
also of $U_{sa}(t)$, to realize the evolution   
$\ket{\Psi (t)}$.   
\bibitem{lindblad76} G. Lindblad,   
Comm. Math. Phys. {\bf 48}, 119 (1976). 
\end{thebibliography}
\end{document}